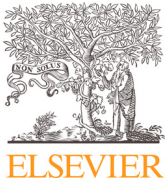
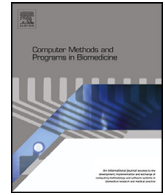

# ISA-Net: Improved spatial attention network for PET-CT tumor segmentation

Zhengyong Huang [a,b,1], Sijuan Zou [c,1], Guoshuai Wang [a,b], Zixiang Chen [a,d], Hao Shen [a,b], Haiyan Wang [a,b], Na Zhang [a,d], Lu Zhang [e,f], Fan Yang [e,f], Haining Wang [g], Dong Liang [a,d], Tianye Niu [h], Xiaohua Zhu [c,*], Zhanli Hu [a,d,*]

[a] *Lauterbur Research Center for Biomedical Imaging, Shenzhen Institute of Advanced Technology, Chinese Academy of Sciences, Shenzhen, 518055, China*
[b] *University of Chinese Academy of Sciences, Beijing, 101408, China*
[c] *Department of Nuclear Medicine and PET, Tongji Hospital, Tongji Medical College, Huazhong University of Science and Technology, Wuhan, 430000, China*
[d] *Chinese Academy of Sciences Key Laboratory of Health Informatics, Shenzhen, 518055, China*
[e] *Brain Cognition and Brain Disease Institute (BCBDI), Shenzhen Institute of Advanced Technology, Chinese Academy of Sciences, Shenzhen, 518055, China*
[f] *Shenzhen-Hong Kong Institute of Brain Science-Shenzhen Fundamental Research Institutions,Shenzhen, 518055, China*
[g] *United Imaging Research Institute of Innovative Medical Equipment, Shenzhen, 518045, China*
[h] *Institute of Biomedical Engineering, Shenzhen Bay Laboratory, Shenzhen, 518118, China*

**ABSTRACT**

Background and Objective: Achieving accurate and automated tumor segmentation plays an important role in both clinical practice and radiomics research. Segmentation in medicine is now often performed manually by experts, which is a laborious, expensive and error-prone task. Manual annotation relies heavily on the experience and knowledge of these experts. In addition, there is much intra- and interobserver variation. Therefore, it is of great significance to develop a method that can automatically segment tumor target regions. Methods: In this paper, we propose a deep learning segmentation method based on multimodal positron emission tomography-computed tomography (PET-CT), which combines the high sensitivity of PET and the precise anatomical information of CT. We design an improved spatial attention network(ISA-Net) to increase the accuracy of PET or CT in detecting tumors, which uses multi-scale convolution operation to extract feature information and can highlight the tumor region location information and suppress the non-tumor region location information. In addition, our network uses dual-channel inputs in the coding stage and fuses them in the decoding stage, which can take advantage of the differences and complementarities between PET and CT. Results: We validated the proposed ISA-Net method on two clinical datasets, a soft tissue sarcoma(STS) and a head and neck tumor(HECKTOR) dataset, and compared with other attention methods for tumor segmentation. The DSC score of 0.8378 on STS dataset and 0.8076 on HECKTOR dataset show that ISA-Net method achieves better segmentation performance and has better generalization. Conclusions: The method proposed in this paper is based on multi-modal medical image tumor segmentation, which can effectively utilize the difference and complementarity of different modes. The method can also be applied to other multi-modal data or single-modal data by proper adjustment.

© 2022 Elsevier B.V. All rights reserved.

## 1. Introduction

Tumor segmentation, depicting a tumor region in patients with positron emission tomography (PET), computed tomography (CT) or magnetic resonance imaging (MRI), is a fundamental task in medical image analysis. It is widely used in many clinical applications, including disease diagnosis, radiomics analysis, treatment planning, personalized medicine and treatment delivery[1]. CT imaging, which was first applied to medical analysis, focuses on the anatomical information of objects and is often used to examine the chest and abdomen of tumor patients. CT examination of the chest shows clearer structures and is more sensitive than conventional X-ray chest films in detecting and showing the accuracy of lesions in the chest; in particular, for the confirmation of early lung cancer, CT of the chest has decisive significance. However, CT examinations use X-ray transmission, so there is a certain amount of radiation delivered to the human body, and the clarity of

* Corresponding author.
*E-mail address:* zl.hu@siat.ac.cn (Z. Hu).
[1] These authors contributed equally.

CT imaging of soft tissue is not sufficient. PET is a novel molecular imaging technique that uses radioactive tracers to display biomolecular metabolism and receptor and neuromediator activity in living subjects, and it can be used for quantitative imaging of physiological, biochemical and pharmacological processes, including blood flow, metabolism, receptors, enzymes and markers themselves [2]. Thus, PET has extensive clinical and research applications in oncology, cardiology and neurology. $^{18}$F-fluorodeoxyglucose ($^{18}$F-FDG) is currently the most commonly used radiotracer. In general, cells in the tumor region are highly metabolized; therefore, the tracer uptake concentration in this region is higher. We can perform a semiquantitative analysis of $^{18}$F-FDG uptake using standard uptake values (SUVs) [3], defined as the ratio of the tracer concentration in the region of interest (ROI) to the whole-body concentration [4]. Although PET is effective in early tumor diagnosis, the resolution of PET images is poor compared to that of CT images. In addition, areas such as the brain and heart generally show high tracer uptake due to high metabolism, which is not good for determining tumor boundaries. Therefore, PET is generally not used as the only means of tumor diagnosis.

The integrated imaging modality PET-CT combines PET and CT so that the advantages of the two imaging technologies complement each other. PET images provide molecular information such as function and metabolism information, while CT provides fine anatomical and pathological information. Through this fusion technology, pathophysiological changes and morphological changes in diseases can be obtained in one image, and this is more suitable for visual perception, especially for radiologists. Fused images reduce uncertainty and minimize the redundant information in the output while maximizing the relevant information. In the past 10 years, PET-CT has increased the correct diagnosis rate of tumors from approximately 85% to 95% ∼ 99%.

Compared with natural images, medical images have distinctive characteristics, such as low resolution, low contrast and scattered targets, and there are also higher requirements for the accuracy and stability of the segmentation algorithm results. Incorrect or unstable segmentation will directly affect the subsequent diagnosis and treatment of patients, thus losing the original purpose of medical image segmentation.

Early image segmentation algorithms were based on traditional methods, such as edge detection, threshold segmentation, filters and other mathematical methods.

Many segmentation methods have been proposed for PET-CT scans. The most intuitive and simple are threshold-based segmentation algorithms, such as the fixed threshold method [5] the adaptive threshold method [6], the iterative threshold method [7] and the histogram analysis method [8], which mainly separates the tumor from the background based on the high contrast between the target and the background by using the SUV of PET or the calculated value of the anatomical information of CT. The SUV is usually used clinically to differentiate malignant tumors from benign lesions and to determine the malignancy of tumors based on the magnitude of the SUV; most scholars consider SUV = 2.5 to be the threshold for differentiating benign from malignant tumors, and SUV > 2.5 is considered to indicate a malignant tumor. However, in practice, the calculation of the SUV is often influenced by several factors, such as patient weight, blood glucose concentration, PET scan time, image reconstruction method and environmental noise. Among these, weight has a significant impact on the SUV, especially for patients with obesity. For patients with diabetes, the SUV calculation is also prone to abnormalities. In addition, the target boundary can be blurred due to a series of factors, such as the uncertainty of the pathology itself, the lack of clarity of the PET image and patient motion. All of these factors make it difficult to achieve good results with the threshold-based segmentation method.

Various deep learning-based methods have also been applied to medical image segmentation, and all of them have achieved good results. The U-Net network proposed by Ronneberger et al. [9] and its variant networks are now the most mainstream means of medical image segmentation; they fuse feature maps of different stages by skip connections, allowing the network to propagate contextual information to higher resolutions, thus achieving a better segmentation effect. Additionally, Milletari et al. [10] proposed the V-Net network, which is another 3D implementation of U-Net. Since then, many U-Net-based network structures have been proposed, such as U-Net++ [11], ResU-Net [12], BCDU-Net [13], and R2U-Net [14]. Nevertheless, most of the studies have been implemented based on a single modality, such as PET, CT or MRI alone [15–17].

There are also many approaches [18–20] based on PET-CT multimodal tumor segmentation, most of which lie in how to better extract the feature information of various modalities and how to fuse the feature information of different modalities. Ashnil et al. [21] proposed a co-learning approach on how to fuse the information of PET-CT in the input stage. Li et al. [22] proposed a deep learning-based variational approach for automatic fusion of the multimodal information of tumors in PET-CT. Zhong et al. [23] proposed a 3D fully convolutional network for combined PET-CT segmentation, which first generated an FCN probabilistic graphical model using U-Net and then performed segmentation based on the probabilistic graphical model. Fu et al. [24] proposed a multimodal spatial attention method, which fuses the PET-generated spatial attention maps weighted to fuse into CT-generated feature information. Bi et al. [25] proposed a cyclic fusion network that fuses features of complementary multimodal images with intermediate segmentation results at each stage, thus reducing the risk of inconsistent feature learning.

The attention mechanism was first proposed in the field of visual images, and DeepMind et al. [26] successfully used the attention mechanism in RNN models for image classification in 2014. Subsequently, Bahdanau et al. [27] successfully applied the attention mechanism to the field of natural language processing (NLP). In 2017, the Google machine translation team proposed the Transformer structural model [28], and the attention mechanism was a focus once again. Currently, attention mechanisms based on CNN models have become a popular research topic. For example, Hu et al. [29] proposed the squeeze-and-excitation network (SE-Net) based on the channel attention mechanism in 2017, and Woo et al. [30] proposed convolutional block attention module(CBAM) in 2018, which is capable of attentional learning on both the channel and spatial dimensions. In addition, X. Li et al. [31] proposed the selection kernel network (SK-Net) in 2019.

Regarding the segmentation of head and neck tumors, Iantsen et al. [32] proposed a model based on a U-Net architecture with residual layers and supplemented with 'Squeeze and Excitation' (SE) normalization. Xie et al. [33] proposed a 3D scSE nnU-Net model, improving upon the 3D nnU-Net by integrating the spatial and channel 'Squeeze and Excitation' (scSE) blocks. They trained the model with a weighted combination of Dice and cross-entropy losses. And in 2021, they used a well-tuned patch-based 3D nnU-Net [34] with standard preprocessing and training scheme, where the learning rate is adjusted dynamically using polyLR. The SE normalization was also one of the main ingredient of their approach [35]. The approach is straighforward yet efficient as they ranked first for HECKTOR Segmentation Challenge 2021. An et al. [36] proposed a framework which is based on the subsequent application of three different U-Nets. The first U-Net is used to coarsely segment the tumor and then select a bounding box. Then, the second network performs a finer segmentation on the smaller bounding box. Finally, the last network takes as input the concatenation of PET, CT and the previous segmentation to refine the predictions.

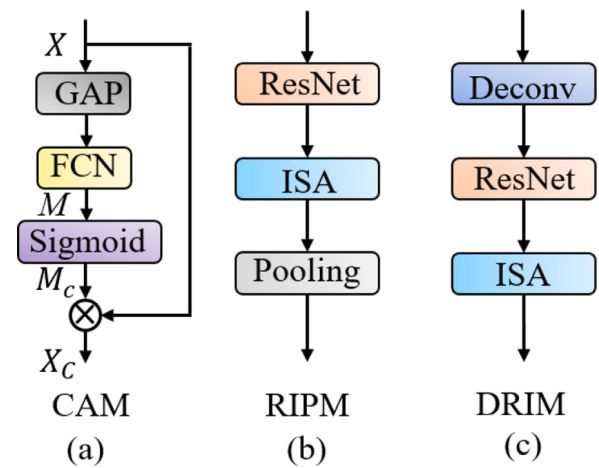

Fig. 2. Some sub-modules of the entire network. (a): Channel attention module and GAP and FCN are abbreviations for global average pooling and fully connected network respectively; (b): Arithmetic module in the encoding stage; (c): Arithmetic module in the decoding stage.

Motivated by these works, we propose an improved spatial attention tumor segmentation network based on combined PET-CT dual modal data, which is designed as a separate sub-module to implement self-attentive learning. We use multi-scale convolution to extract feature information, and we design an improved spatial attention (ISA) network to increase the sensitivity of PET or CT in detecting tumors, which can highlight the tumor region location information and suppress the non-tumor region location information. Finally, we validate the superiority of the algorithm on two different datasets.

The rest of this paper is organized as follows: Section 2 introduces the proposed method. Section 3 describes our experimental method and the results on two datasets in detail. Section 4 analyzes the results and the existing deficiencies. Section 5 summarizes the full paper.

## 2. Methods

### 2.1. Residual module

The residual network (ResNet) was proposed in 2016 [37], and compared to the previous convolutional networks, it adds a shortcut connection, as shown in Fig. 1. The presence of shortcut connections ensures that we can increase the number of layers in the convolutional network without degradation of the model, such as gradient disappearance. In this paper, we use two residual blocks on each feature layer of the 3D U-Net to prevent model degradation; the input and output of the first residual block have different numbers of channels, and its shortcut connection branch contains a convolutional operation with a convolutional kernel size of 11. The number of channels before and after the second residual block remains the same, and the shortcut connection branch does not perform task transformation.

$$X_R = \sum \sigma(\mathcal{F}_{norm}(\mathcal{F}_{pool}(Conv_k(U)))), k = 1, 3 \tag{1}$$

where $Conv_k$ denotes the convolution operation with convolution kernel sizes of $3 \times 3 \times 3$ and $1 \times 1 \times 1$, $\mathcal{F}_{pool}$ denotes the max-pooling operation, $\mathcal{F}_{norm}$ denotes the instance normalization operation, $\sigma$ is a nonlinear activation function, and the rectified linear unit (ReLU) is used.

### 2.2. Channel attention module

Since attention mechanisms were first proposed, many structures based on attention mechanisms have been proposed successively; channel attention is a common attention mechanism, such as in SE-Net [29] and SK-Net [31]. As shown in Fig. 2(a), channel attention first obtains a channel attention vector $M \in \mathbb{R}^{C \times 1}$ by performing global average pooling and fully connected layer squeezing on the input feature graph $X \in \mathbb{R}^{D \times H \times W}$; $C$ denotes the number of feature graphs, i.e., the number of feature channels. $M$ then undergoes sigmoid activation to obtain the final channel attention graph, which is multiplied with the original input features $X$ in the channel dimension to yield our desired output feature map $X_C \in \mathbb{R}^{D \times H \times W}$.

$$M_c = \sigma(\mathcal{F}_{fc}(\mathcal{F}_{GAP}(X))) \tag{2}$$

$$X_c = M_c * X \tag{3}$$

where $\mathcal{F}_{GAP}$ denotes the global average pooling operation, $*$ denotes the elementwise product, and $\mathcal{F}_{fc}$ is two fully connected layers, where the first layer contains $2C$ neuron nodes and the second layer contains $C$ neuron nodes; $\sigma$ is the nonlinear activation function, and sigmoid is used in this paper.

### 2.3. Improved spatial attention module

The structure of the improved spatial attention network (ISA-Net) is shown in Fig. 3. We first extract the desired feature information based on a global channel attention network and then divide each obtained feature map $X \in \mathbb{R}^{D \times H \times W}$ into 3 branches; each branch performs the convolution operation separately to obtain the corresponding feature information $X_i$:

$$X_i = K_i \otimes X, i = q, k, v \tag{4}$$

where $\otimes$ and $K_i$ denote convolution operations and different branches of the convolution operations, respectively, and all $K_i$ have the same number of convolution kernels. Then, we fuse branch $q$ and branch $k$ as the weight information:

$$U = X_q \cdot X_k \tag{5}$$

where $U \in \mathbb{R}^{D \times H \times W}$ and $\cdot$ denotes the matrix inner product. After obtaining the fused feature information, we flatten each feature map into a one-dimensional vector $S \in \mathbb{R}^{L \times 1}$, where $L = D * H * W$.

$$S = \mathcal{F}_F(U) \tag{6}$$

$S$ is a one-dimensional vector connected by all the pixel points of $U$, and $\mathcal{F}_F$ represents the conversion of the three-dimensional input $U$ into the one-dimensional vector $S$. Each pixel value in the feature map represents the importance of the current location

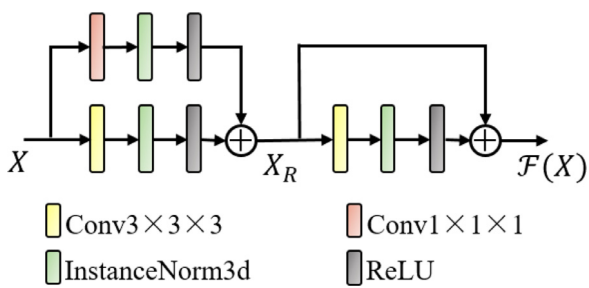

Fig. 1. The shortcut connection of the first residual block contains a convolution operation with a convolution kernel size of $1 \times 1 \times 1$ to change the number of convolution channels; that is, $X$ and $X_R$ have different numbers of channels. The second residual block's shortcut connection has no transformation, and $X_R$ and $\mathcal{F}_X$ have the same number of channels.

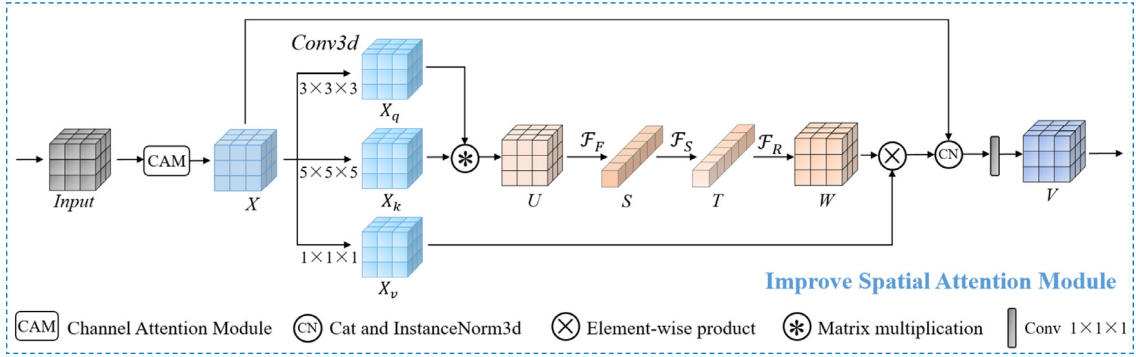

**Fig. 3.** Improved spatial attention module. The input is the output of the previous residual module, and the output of the network has the same dimension as the input.

information, and $S$ is transformed by softmax to obtain the weight of each pixel location.

$$T = \mathcal{F}_S(S) \tag{7}$$

Each element of the vector $S$ represents the importance of a pixel in the original feature map, and $\mathcal{F}_S$ indicates softmax transformation. To add the learned weight information to the original feature map, $T$ is reconstructed as a weight matrix $W \in \mathbb{R}^{D \times H \times W}$ with the same dimensional information as the input $X$. $\mathcal{F}_S$ represents the conversion of the one-dimensional input $T$ into the three-dimensional matrix $W$.

$$W = \mathcal{F}_R(T) \tag{8}$$

The activation function used for all the convolution operations in the ISA-Net module is ReLU.

### 2.4. Overall architecture

Our model takes two 3D U-Nets as the baseline; the main components are the two main channels, the CT channel and PET channel, and these two channels are symmetrical to each other. The network contains two main modules, the ResNet module and the ISA-Net module. For each individual CT or PET channel, we designed a total of 5 layers of the 3D U-Net structure. Each layer consists of a ResNet module and an ISA-Net module, where the ResNet module is mainly used for extracting deep-level features and preventing model degradation; another role of it is to increase the number of convolutional kernels. The ISA-Net module is located after the ResNet module, which is a local attention module that is conducive to finding the location information of lesions. The overall structure of our model is shown in Fig. 4.

### 2.5. Loss function

In the image segmentation task, we focus on predicting the similarity between samples and tokens. In general, we use the Dice similarity coefficient (DSC) to evaluate the segmentation results. To maximize the DSC, we use the Dice loss function [10] in the training process. The Dice loss function is described as follows:

$$Loss_D(y, \hat{y}) = 1 - \frac{2 \sum_i^N y_i \hat{y}_i}{\sum_i^N y_i + \sum_i^N \hat{y}_i + \epsilon} \tag{9}$$

where $y$ and $\hat{y}$ are the label and prediction result of the network, respectively, $N$ is the total number of pixels and $\epsilon$ is a small constant to prevent division by zero. In the training process, to address individual bad data, we also use the focal loss function [38], which can solve the model training problem caused by sample nonequilibrium when the samples are difficult to classify:

$$Loss_F(y, \hat{y}) = -(1-\alpha)(1-y)\hat{y}^\gamma \log(1-\hat{y}) - \alpha y (1-\hat{y})^\gamma \log \hat{y} \tag{10}$$

where $\alpha$ is the balanced weight factor and the parameter $\alpha$ is set to 0.5. $\gamma$ is the rate of weight decline, and $\gamma$ is set to 2.

Ultimately, we combine these two components as the loss function during network training.

$$Loss_{Total}(y, \hat{y}) = Loss_D(y, \hat{y}) + Loss_F(y, \hat{y}) \tag{11}$$

### 2.6. Evaluation metrics

#### 2.6.1. Dice similarity coefficient

The DSC is the most frequently used metric in medical image segmentation, and it is a set similarity metric that is usually used to calculate the similarity of two samples with a value domain of [0, 1]. The best segmentation result is 1, and the worst result is 0. In this paper, we use the DSC as the main evaluation metric and save the best training model based on the highest DSC score during the training process. Its calculation method is as follows:

$$Dice = 2 * \frac{pred \cap true}{pred \cup true} \tag{12}$$

where $pred$ is the set of voxels of the predicted values and $true$ is the set of voxels of the true values.

#### 2.6.2. Hausdorff distance

The Hausdorff distance (HD) is a shape similarity metric that can be understood as the maximum value of the shortest distance from an element in one set to an element in another set. It focuses on the segmentation boundary and can be used as a complement to the DSC, which focuses on the interior of the ROI while the Hausdorff distance focuses more on the information of the edge positions.

$$\begin{aligned} HD(A, B) &= \max(h(A, B), h(B, A)) \\ h(A, B) &= \max_{a \in A} \{\min_{b \in B} ||a - b||\} \\ h(B, A) &= \max_{b \in B} \{\min_{a \in A} ||b - a||\} \end{aligned} \tag{13}$$

where $||\cdot||$ is the distance norm of set $A$ and set $B$, such as the $L2$ norm or the Eulerian distance norm, and the latter is used in this paper.

#### 2.6.3. Average symmetric surface distance

The average symmetric surface distance (ASSD) is calculated as the average of the distances of all surface points between two voxels and is also an evaluation metric that focuses on segmented edges.

$$ASD(X, Y) = \sum_{x \in X} \min_{y \in Y} d(x, y) / X \tag{14}$$

$$ASSD(X, Y) = \{ASD(X, Y) + ASD(Y, X)\}/2 \tag{15}$$

where $X$ and $Y$ denote the two voxel sets and ASD is the average surface distance of two voxels.

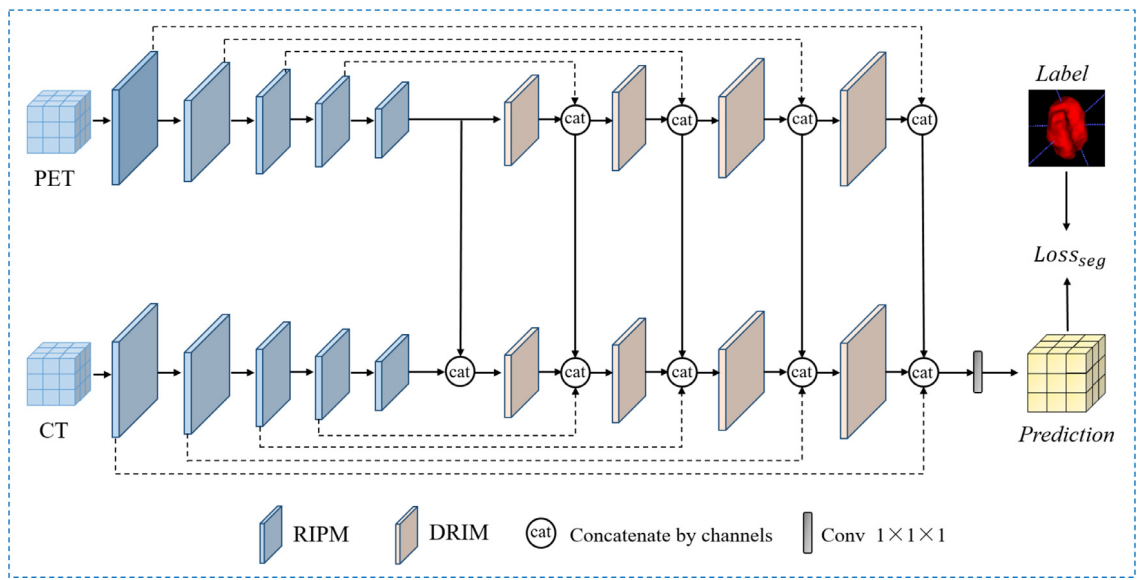

**Fig. 4.** Overall network structure. The network contains two U-Nets and their weights are not shared, and the input is three-dimensional images with size of 144 × 144 × 144. PET images and CT images are input into the network independently, and the initial number of convolution kernels in both branches is 16. After encoding, the features obtained from PET and CT are fused by connecting them according to the channel dimensions. In the fusion process, the feature maps of the two modalities occupy the same weight, i.e., the two are fused in the ratio of 1:1. The number of convolution kernels is doubled after each RIPM processing step and halved after each DRIM processing step.

There are many other evaluation indicators, such as the volume overlap error (VOE), relative volume difference (RVD), recall and precision.

## 3. Experiments and results

### 3.1. Datasets

To validate the proposed method, we use two datasets for testing, the head and neck tumor (HECKTOR) dataset[39,40] and the soft tissue sarcomas (STS) dataset[41]. The HECKTOR dataset contains a total of 224 cases from five different centers with the same scanning protocol, each containing CT and PET images and a primary gross tumor volume (GTV), where the original image resolutions were 512×512×91 at 0.977 mm×0.977 mm×3.270 mm for CT and 128×128×91 at 3.516 mm×3.516 mm×3.270 mm for PET. The STS dataset includes 51 patients with extremity sarcomas, with data from different sites and scanners. Four different imaging modalities were obtained for each patient: two paired MRI (T1 and T2) scans and one PET/CT scan. MRI and PET/CT examinations were obtained on different days, which resulted in changes in body position as well as anatomical variations. In this paper, we design a segmentation network based on PET/CT scan results. For each patient, there are CT and PET images and primary GTV. The original image resolutions were 512×512×267 at 0.977 mm×0.977 mm×3.270 mm for CT and 128×128×267 at 3.906 mm×3.906 mm×3.270 mm for PET.

Before the experiment, we preprocessed these two datasets, including performing data conversion and resampling. First, we converted the intensity values of the CT images and PET images into Hounsfield units and SUVs, respectively. Then, we sampled the data to a resolution of 1 mm×1 mm×1 mm and a size of 144 × 144 × 144 by using the trilinear interpolation.

### 3.2. Experimental details

All models were trained using an NVIDIA RTX 3090 GPU, and all hyperparameter settings used were kept consistent. The networks were trained end-to-end for 300 epochs. Considering the GPU memory limitation, the batch size was set to 1. The model used the Adam optimizer; the first momentum factor was set to 0.9, and the second momentum factor was set to 0.99. The initial learning rate was $3 \times 10^{-4}$, and the learning rate was updated using a simulated annealing algorithm, which decayed by $1 \times 10^{-6}$ after every 25 epochs and then reinitialized the learning rate to 0.0003, which can prevent falling into local suboptimal solutions to some extent. Before the model began training, the weights of the network filter kernel were initialized using the method of He et al.[42], and the filter kernel did not use bias terms.

All inputs were subject to normalization. The Hounsfield units of CT images in the range of [-1024,1024] are normalized to the range of [-1,1], and the PET images were normalized using the mean and standard deviation:

$$x'_{ct} = \frac{x_{ct}}{\max(x_{ct})} \quad (16)$$

$$x'_{pet} = \frac{1}{\sigma}(x_{pet} - \mu) \quad (17)$$

where $\mu = E(x_{pet})$, $\sigma = \sqrt{Var(x_{pet}) + \epsilon}$, and $\epsilon$ is a small constant used to prevent division by zero.

In addition, we made data enhancements to the training set, including random rotation (probability of rotation: 0.5, angular range of rotations: $0° \sim 45°$), mirroring (probability of mirroring: 0.5), and the mixup [43] data enhancement strategy (probability of use is 0.2), which can improve the generalization capability of the network. The so-called mixup data enhancement strategy randomly selects two training samples and fuses them according to a certain proportion. It needs to fuse not only the input image but also the corresponding label.

$$\begin{aligned} f &= \lambda f_1 + (1-\lambda)f_2 \\ g &= \lambda g_1 + (1-\lambda)g_2 \end{aligned} \quad (18)$$

where $f$ represents the input image, which includes a CT image and a PET image. $g$ indicates the corresponding ground truth. In this article, we set $\lambda = 0.5$, and the probability of enhancement with mixup data is 0.2. An example of mixup data enhancement is shown in Fig. 5.

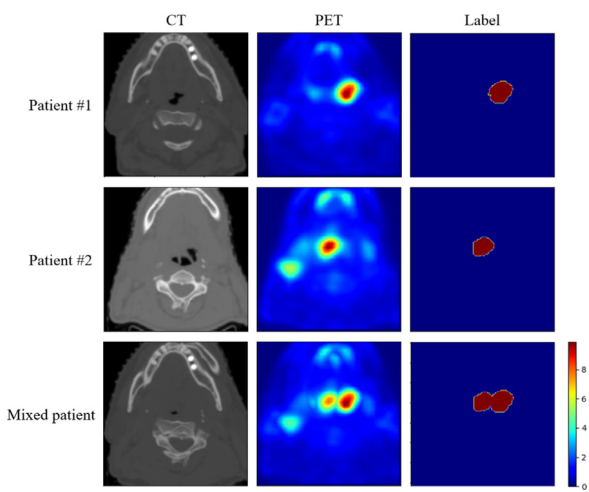

**Fig. 5.** An example of mixup data enhancement.

**Table 1**
Results for different input modalities based on our method.

| Dataset | Modality | DSC | HD | ASSD |
|---|---|---|---|---|
| STS | CT | 0.7331 | 21.8172 | 5.4372 |
|  | PET | 0.8080 | 18.1022 | 4.8839 |
|  | PET+CT | **0.8378** | **15.3266** | **4.5434** |
| HECKTOR | CT | 0.6434 | 6.2747 | 2.4888 |
|  | PET | 0.7356 | 4.8412 | 1.8414 |
|  | PET+CT | **0.8076** | **3.4615** | **1.2099** |

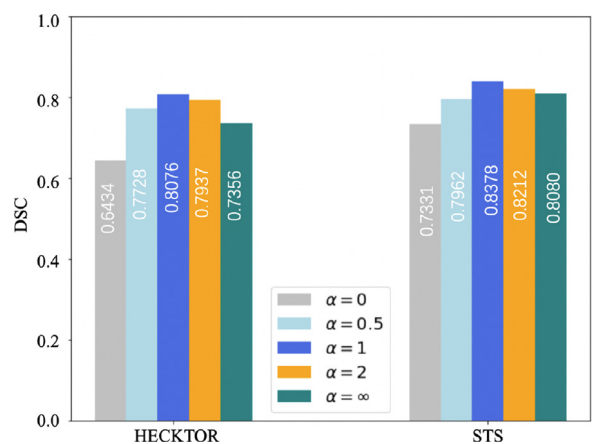

**Fig. 6.** The results of different weighting factor assigned to each modality during fusion. $\alpha = 0$ means only CT data are used, and $\alpha = \infty$ means only PET data are used.

Due to the small amount of data, we used five-fold cross-validation for both datasets and all methods and averaged the results as the final evaluation results. The datasets were randomly divided into a training set and a test set, which were 80% and 20% of the data respectively, and the same data division was used for each method.

### 3.3. Results for different input modalities based on our method

Table 1 shows the segmentation results of the proposed network for both datasets with different input combinations. When the input data is only PET or CT, the network structure is the CT branch in Fig. 4, i.e., the PET branch in Fig. 4 is cropped out. It can be seen from the results of both datasets that the segmentation using CT images alone is the worst, the segmentation of PET images is significantly better, and the best results are obtained using PET/CT dual-modality images as input with the highest DSC score. This result is expected, as we know that PET is a molecular imaging technique that is based on the physiological activity of radiotracer imaging, so it is better for imaging soft tissue such as tumors, and its imaging results can better show the tumor location information, which is helpful for segmentation. In our model, the combined PET/CT dual modality is used as the input information; the CT images can provide accurate anatomical and pathological information, while the PET images can provide edge location information, so the advantages of both imaging techniques complement each other to achieve better segmentation results. In the decoding stage, our network needs to fuse the information of different modal features, so we analyzed the weighting factor assigned to each modality during fusion, and the results are shown in Fig. 6. Where $\alpha$ is a scaling factor defined as $\alpha = w_p/w_c$, $w_p$ and $w_c$ denote the weights of PET and CT during feature fusion, respectively. In the experiments, we mainly analyzed five five different fusion ratios, namely $\alpha \in [0, 0.5, 1, 2, \infty]$. $\alpha = 0$ means only CT data are used, and $\alpha = \infty$ means only PET data are used. The results show that the best results can be achieved when PET and CT are fused in a 1 : 1 ratio. All of our subsequent experiments were conducted according to this ratio of fusion.

### 3.4. Qualitative comparison with other methods

As seen from Table 1, better results can be achieved using PET-CT dual-modality input, so our network is also designed based on PET-CT dual-modality input. The data from two different modalities were input into different channels. Fig. 7 shows the segmentation results for different views of a sample in the STS dataset, and it can be seen that only our method correctly segmented some of the nontumor locations that exhibit high uptake regions (with high SUVs) in the PET image. To better illustrate the superiority of the ISA-Net module, Fig. 8(a) shows the segmentation borders of a sample of the STS dataset, and Fig. 8(b) shows the segmentation borders of a sample of the HECKTOR dataset, and we can see that our method has the best segmentation on this slice, while the other algorithms show oversegmentation and fail to capture more subtle edge information. The segmentation results of our model overlap better with the real ROI, which further illustrates that our model can better capture the tumor edge information.

### 3.5. Quantitative comparison with other methods

The above qualitative analysis shows the superiority of our model, but it is still not completely convincing. Therefore, we analyzed the prediction results of our model with some quantitative metrics, such as the DSC, HD, and precision. Table 2 clearly shows the computed results of the quantitative metrics of our proposed method and other methods, where the training strategy and hyperparameters are kept the same for all methods. From the results, it is clear that our segmentation method achieves the best results for all evaluation metrics. All methods use the same model structure, i.e. PET and CT are input from two separate channels. The only difference between them is the replacement of the ISA module with the corresponding SE, SK or CBAM module.

### 3.6. Compare with state-of-the-art methods

In this section, we compared our method with the part of methods of the HECKTOR Segmentation Challenge 2020[44] and 2021[45] on two datasets and Table 3 shows the comparison results. Among the four comparison algorithms, the implementation code of Iantsen et al.'s method can be found at https://github.com/iantsen/hecktor. There is no open source code for the other three methods, and we reproduce them strictly according to the description of the paper. In the experiments, the hyperparameter settings of all algorithms were kept the same, where [35] used SGD optimizer with an initial learning rate of 0.001, and the remaining three algorithms used Adam optimizer with an initial learning rate of 0.0001. In both DSC and HD and ASSD metrics, our results are far better than these four comparison methods, especially in DSC metrics, our method has a great improvement. In addition, the improvements in HD and ASSD also reflect the superiority of our method in tumor margin processing. These results show that our proposed ISA-Net does improve the accuracy of tumor segmentation, which can not only find the location of tumors, but also accurately extract the edge information of tumors to achieve higher accuracy of segmentation.

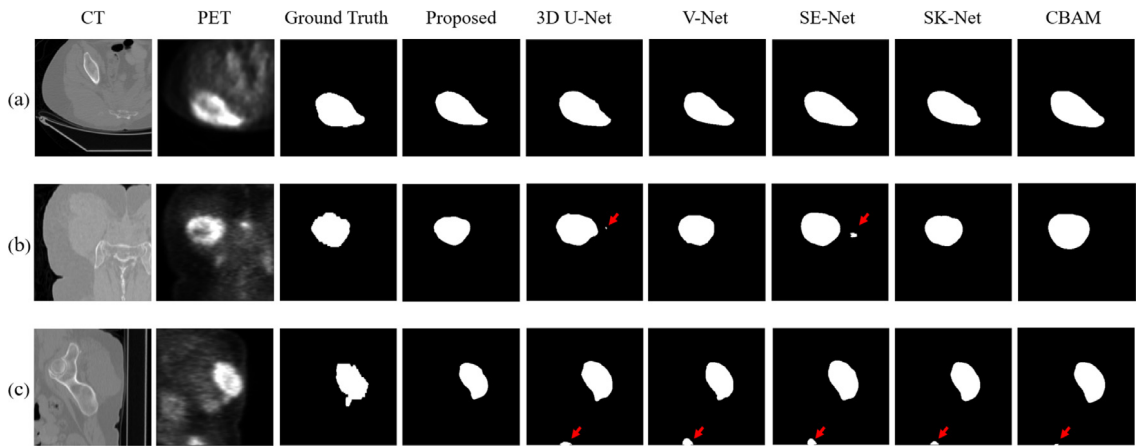

**Fig. 7.** Segmentation results from different views of one sample in the STS dataset. (a): Axial view; (b): Coronal view; (c): Sagittal view.

Table 2
Quantitative comparison of the segmentation performance of different methods with manual segmentation as the reference (mean). Note: DSC - Dice similarity coefficient, HD - Hausdorff distance, VOE - Volume overlap error, RVD - Relative volume difference, ASSD - Average symmetric surface distance.

| Dataset | Method | DSC | HD | ASSD | RVD | VOE | Recall | Precision |
|---|---|---|---|---|---|---|---|---|
| STS | U-Net [9] | 0.8132 | 22.8454 | 4.8052 | 0.2194 | 0.1404 | 0.7873 | 0.8497 |
| | V-Net [10] | 0.7769 | 25.0871 | 6.3327 | 0.2701 | 0.1831 | 0.7401 | 0.8557 |
| | U-Net+SE-Net [29] | 0.8270 | 21.9796 | 4.8852 | 0.2427 | 0.1149 | 0.8096 | 0.8312 |
| | U-Net+SK-Net [31] | 0.8197 | 19.2891 | 4.7980 | 0.2095 | 0.1092 | 0.7908 | **0.8696** |
| | U-Net+CBAM [30] | 0.8302 | 17.5516 | 4.7655 | 0.2017 | **0.1018** | 0.8121 | 0.8562 |
| | U-Net+ISA-Net | **0.8378** | **15.3266** | **4.5434** | **0.1868** | 0.1207 | **0.8126** | 0.8611 |
| HECKTOR | U-Net [9] | 0.7665 | 8.5091 | 1.3727 | 0.3175 | 0.1815 | 0.8209 | 0.7960 |
| | V-Net [10] | 0.7553 | 7.3114 | 1.7345 | 0.3937 | 0.2189 | 0.8089 | 0.7969 |
| | U-Net+SE-Net [29] | 0.7712 | 6.2971 | 1.4513 | 0.3190 | 0.1753 | 0.8205 | 0.7846 |
| | U-Net+SK-Net [31] | 0.7894 | 3.9534 | 1.4223 | 0.2993 | 0.1663 | 0.8235 | 0.7457 |
| | U-Net+CBAM [30] | 0.7806 | 4.1417 | 1.4216 | 0.2602 | 0.1549 | **0.8364** | 0.7958 |
| | U-Net+ISA-Net | **0.8076** | **3.4615** | **1.2099** | **0.2172** | **0.1427** | 0.8351 | **0.8043** |

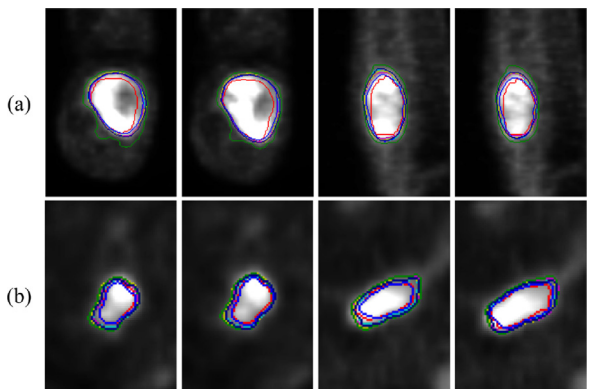

**Fig. 8.** Identification of a tumor border with the proposed ISA-Net (blue contour), CBAM (navy contour), SK-Net (cyan contour), SE-Net (medium purple contour), 3D U-Net (yellow contour) and V-Net (green contour). The red contour indicates the ground truth. The first two columns are the segmentation results of the axial view, and the last two columns are the segmentation results of the sagittal view. (a): A sample from the STS dataset; (b): A sample from the HECKTOR dataset. (For interpretation of the references to colour in this figure legend, the reader is referred to the web version of this article.)

Table 3
Comparison results based on HECKTOR and STS datasets with state-of-the-art methods.

| Dataset | Method | DSC | HD | ASSD |
|---|---|---|---|---|
| STS | Iantsen et al. [32] | 0.8126 | 18.2384 | 6.8501 |
| | Xie et al. [33] | 0.8002 | 19.0614 | 6.6018 |
| | Peng et al. [35] | 0.7639 | 16.8815 | 4.7174 |
| | An et al. [36] | 0.7943 | 15.9484 | 5.2321 |
| | Ours | **0.8378** | **15.3266** | **4.5434** |
| HECKTOR | Iantsen et al. [32] | 0.7721 | 6.4286 | 1.8501 |
| | Xie et al. [33] | 0.7666 | 7.2180 | 1.6018 |
| | Peng et al. [35] | 0.7591 | 5.0749 | 1.4174 |
| | An et al. [36] | 0.7602 | 4.6523 | 1.5321 |
| | Ours | **0.8076** | **3.4615** | **1.2099** |

## 4. Discussion

In this paper, we propose an ISA-Net for tumor segmentation based on combined PET-CT scans, which uses multi-scale convolution to extract feature information. To achieve a better segmentation effect, we also use other techniques to enhance the feature extraction; for example, the channel attention module is used to obtain global feature information, and the residual network is used to deepen the network while preventing model degradation. Re-sults show that our proposed segmentation mothed can highlight the tumor region location information and suppress the non-tumor region location information and is more advantageous than some current attention methods in accuracy and generalization.

*4.1. Result analysis*

We first compare the effect of segmentation using dual-modal information and single-modal information based on the 3D U-Net baseline. Based on the experimental results in Table 1, we design a segmentation network ISA-Net based on dual-modal PET-CT scans. In ISA-Net, a channel attention module is first used to extract the global feature information, after which we design an improved spatial attention network. This network first implements two different convolution operations to extract feature information, and then the features are fused through matrix multiplication. The results are taken as the original input characteristics of the weight information. In our experiments, we found that two convolutional operations with different convolutional kernel sizes (in this paper, we use $3 \times 3 \times 3$ and $5 \times 5 \times 5$ convolutional kernel sizes) work better than operations with the same convolutional kernel size, and we infer that the reason is that the large convolutional kernel size expands the local perceptual field, enabling the extraction of more abundant characteristic information. In addition, we performed a separate convolution operation with a convolution kernel size of 1 and took the result as the main feature information for the original feature input. The main purpose of using a convolution kernel of size $1 \times 1 \times 1$ is to change the feature dimension of the input and to reduce the information loss as much as possible.

As shown in Fig. 4, our network uses two independent channels for the PET and CT information input in the encoding stage, which can extract the feature information of PET and CT images more fully than sharing one channel. In the decoding stage, we fuse the features extracted from the two channels. Both the qualitative and quantitative results show that our proposed self-attentive module is superior to other attention methods in tumor segmentation. And our approach also surpasses current state-of-the-art methods for segmenting head and neck tumors.

In PET images, tumor areas are represented as high-uptake areas (hot spots), which can be used as tumor location information to increase tumor segmentation accuracy. However, some parts of the body also exhibit high-uptake areas (hot spots) due to relatively vigorous metabolic activities. Correctly determining whether these hot spots are benign or malignant is a difficult problem for tumor segmentation. In one patient shown in Fig. 7, the PET image contains two main hot spots, and only our method achieves correct segmentation.

In addition, the datasets that we used to verify the proposed algorithm are different. The volumes of head and neck tumors are relatively small, and the distribution location information is fixed. Soft tissue sarcomas are relatively large in size and complex in distribution. Nevertheless, our algorithm achieves good performance on both datasets, which indicates that our algorithm has good generalization and can be applied in the segmentation of other types of tumors.

*4.2. Limitations and future work*

As shown in Table 2, although our algorithm achieves good results on all evaluation metrics, the HD metric is still larger on the STS dataset, and we infer that the reason for this is the large size of soft tissue sarcomas in morphology, with varying shapes and sizes and uncertain distribution locations in the body. The HD metric focuses on the segmentation boundary, which indicates that our algorithm is deficient in edge detail extraction; this is the direction of our future work. We hope to develop a postprocessing algorithm to optimize the segmentation boundary to achieve better segmentation results.

During the training process, we found that the model was complex and the computational cost was high, which led us to seting the batch size to 1 due to limited memory. Next, we will further optimize the model to improve performance while reducing memory consumption and making the model more lightweight.

## 5. Conclusion

In this paper, we propose an improved spatial attention method based on combined PET-CT multimodal data, which can take full advantage of the high sensitivity feature of PET for tumor detection and high accuracy of CT in tumor boundary structure to extract tumor location information. Finally, we design a two-channel network to validate the proposed method using 3D U-Net as the baseline. Both the qualitative and quantitative experimental results show that our method improves the tumor segmentation results, and the comparative experiments show that our method outperforms some other attention methods. In addition, our method is superior to the state-of-the-art methods for head and neck tumor segmentation.

## Declaration of Competing Interest

The authors declare that they have no conflict of interest.

## Acknowledgments

This work was supported by the National Natural Science Foundation of China (32022042, 81871441, 91959119), the Shenzhen Excellent Technological Innovation Talent Training Project of China (RCJC20200714114436080), the Natural Science Foundation of Guangdong Province in China (2020A1515010733), Chinese Academy of Sciences Key Laboratory of Health Informatics in China (2011DP173-015), the Guangdong Innovation Platform of Translational Research for Cerebrovascular Diseases of China.